\documentclass[twocolumn,showpacs,superscriptaddress,preprintnumbers,floatfix]{revtex4}
\usepackage{graphicx,epsf,amssymb,amsmath,latexsym}
\newcommand{\be}{\begin{eqnarray}}
\newcommand{\ee}{\end{eqnarray}}

\renewcommand{\d}{\mbox{${\rm d}$}}

\begin{document}
\title{Higher Order Slow-Roll Predictions for Inflation}
\author{Roberto Casadio}
\email{Roberto.Casadio@bo.infn.it}
\affiliation{Dipartimento di Fisica, Universit\`a di Bologna and I.N.F.N.,
Sezione di Bologna, via Irnerio~46, 40126 Bologna, Italy}
\author{Fabio Finelli}
\email{finelli@bo.iasf.cnr.it}
\affiliation{IASF/INAF, Istituto di Astrofisica Spaziale e Fisica Cosmica,
Istituto Nazionale di Astrofisica, Sezione di Bologna,
via~Gobetti 101, 40129 Bologna, Italy.}
\author{Mattia Luzzi}
\email{Mattia.Luzzi@bo.infn.it}
\affiliation{Dipartimento di Fisica, Universit\`a di Bologna and I.N.F.N.,
Sezione di Bologna, via Irnerio~46, 40126 Bologna, Italy}
\author{Giovanni Venturi}
\email{armitage@bo.infn.it}
\affiliation{Dipartimento di Fisica, Universit\`a di Bologna and I.N.F.N.,
Sezione di Bologna, via Irnerio~46, 40126 Bologna, Italy}
\begin{abstract}
We study the WKB approximation beyond leading order for cosmological
perturbations during inflation.
To first order in the slow-roll parameters, we show that an
improved WKB approximation leads to {\em analytical\/} results
agreeing to within $0.1\%$ with the standard slow-roll results.
Moreover, the leading WKB approximation to second order in the
slow-roll parameters leads to {\em analytical\/} predictions in
qualitative agreement with those obtained by the Green's function
method.
\end{abstract}
\maketitle
\noindent
{\em Introduction.}
It is nowadays common to state that we are in the era of
precision cosmology.
The implication of present and future data sets will be
able to discriminate among different inflationary
models~\cite{infla}.
For this reason, the comparison of inflationary models with
observations requires theoretical advances in the
predictions for the power spectrum of primordial perturbations
beyond the lowest order in the slow-roll parameters worked
out by Stewart and Lyth~\cite{SL}.
Such slow-roll parameters quantify the deviation from an
exactly exponential expansion during inflation and are
related for a canonical inflaton to the derivative of the
potential.
Among the many different conventions, we find it
convenient to use the hierarchy of horizon flow functions
$\epsilon_i$~\cite{terrero}, defined as
$\epsilon_{i+1}=\dot{\epsilon}_i/\left(H\,\epsilon_i\right)$,
with $\epsilon_1=-\,\dot{H}/H^2$, $H=\dot{a}/a$ the
Hubble parameter, $a$ the scale factor and dots denote
derivatives respect to the cosmic time.
\par
The search for deviations from a simple power-law parameterization
of the primordial power spectrum in the first year of WMAP data 
has begun~\cite{peiris}, with however no statistically
conclusive evidence for a significant deviation.
Stronger conclusions about any deviations will be made possible
by the better resolution of the {\sc Planck\/} satellite~\cite{planck}.
In the context of inflation, the deviations from power-law
spectra considered so far correspond to some of the predictions
beyond the first order in the slow-roll parameters and are
expected in general.
\par
The purpose of this {\em letter\/} is to show how the primordial
power spectrum of cosmological perturbations generated in a single 
field inflationary model can be
{\em analytically\/} predicted by the WKB method with sufficient
accuracy.
We shall also show how two different analytical approximations,
such as the Green's function method
(GFM henceforth)~\cite{gongstewart} and the WKB
method~\cite{MS,WKB1} used here, agree to second order
on the polynomial structure of the results in the horizon flow
functions during inflation, but differ in the numerical
coefficients of the second order terms
(the same polynomial structure in the spectral indices
is found by the uniform approximation~\cite{HHJMP,HHHJMP,LA}).
\par\noindent
{\em Cosmological~perturbations.}
Let us begin by recalling that scalar (density) and tensor
(gravitational wave) fluctuations on a Robertson-Walker
background are given respectively by $\mu=\mu_{\rm S}\equiv a\,Q$
($Q$ is the Mukhanov variable~\cite{mukh}) and
$\mu=\mu_{\rm T}\equiv a\,h$ ($h$ is the amplitude of the two
polarizations of gravitational waves~\cite{gris,staro}).
The functions $\mu$ must satisfy the one-dimensional
Schr\"odinger-like equation
\be
\left[\frac{{\d}^2}{{\d}\eta^2}+\Omega^2(k,\eta)\right]
\,\mu=0
\ ,
\label{osci}
\ee
together with the initial condition
\be
\lim_{\frac{k}{a\,H}\rightarrow +\infty} \mu(k,\eta)
\simeq\frac{{\rm e}^{-i\,k\,\eta}}{\sqrt{2\,k}}
\ .
\label{init_cond_on_mu}
\ee
In the above $\eta$ is the conformal time, $k$ is the wave-number,
and
\be
\Omega^2(k,\eta)\equiv k^2-\frac{z''}{z}
\ ,
\label{freq}
\ee
where $z=z_{\rm S}\equiv a^2\,\phi'/H$ for scalar
and $z=z_{\rm T}\equiv a$ for tensor perturbations
($\phi$ is the homogenous inflaton and primes denote
derivatives with respect to $\eta$).
The dimensionless power spectra of scalar and tensor
fluctuations are then given by
\begin{subequations}
\be
\mathcal{P}_{\zeta}\equiv
\displaystyle\frac{k^{3}}{2\,\pi^{2}}\,
\left|\frac{\mu_{\rm S}}{z_{\rm S}}\right|^{2}
\ ,
\ \ \ \
\mathcal{P}_{h}\equiv
\displaystyle\frac{4\,k^{3}}{\pi^{2}}\,
\left|\frac{\mu_{\rm T}}{z_{\rm T}}\right|^{2}
\label{spectra_def}
\ee
and the spectral indices and runnings by
\be
&&
n_{\rm S}-1\equiv
\left.\displaystyle\frac{\d\ln \mathcal{P}_{\zeta}}
{\d\ln k}\right|_{k=k_{*}}
\ ,
\ \ \
n_{\rm T}\equiv
\left.\displaystyle\frac{\d\ln \mathcal{P}_{h}}
{\d\ln k}\right|_{k=k_{*}}
\label{n_def}
\\
&&
\alpha_{\rm S}\equiv\left.
\frac{\d^{2}\ln\mathcal{P}_{\zeta}}
{(\d\ln k)^{2}}\right|_{k=k_{*}}
\ ,
\ \ \
\alpha_{\rm T}\equiv\left.
\frac{\d^{2}\ln \mathcal{P}_{h}}
{(\d\ln k)^{2}}\right|_{k=k_{*}}
\label{alpha_def}
\ee
where $k_*$ is an arbitrary pivot scale.
We also define the tensor-to-scalar ratio
\be
R\equiv\left.\frac{\mathcal{P}_{h}}{\mathcal{P}_{\zeta}}
\right|_{k=k_{*}}
\ .
\label{R_def}
\ee
\end{subequations}
\par
\noindent
{\em Slow-roll and WKB~approximations.}
One of the very few cases for which the equations for cosmological
perturbations can be integrated exactly is that of
power-law inflation~\cite{PL,LS}, where the inflaton is a canonical
scalar field with an exponential potential.
In such a case, $\epsilon_1$ is constant in time and the hierarchy
of the horizon flow functions is therefore truncated with
$\epsilon_i=0$ for $i\ge 2$.
The original slow-roll approximation corresponds to considering
both $\epsilon_1$ and $\epsilon_2$ constant in time for {\em any\/}
potential~\cite{SL}.
\par
Although apparently inappropriate for studying physical problems
such as the hydrogen atom~\cite{langer} (or cosmological
perturbations~\cite{MS}), the WKB method can be cleverly applied
{\em after\/} suitable redefinitions of the wave-function
(for Fourier modes $k$) and variables (with the corresponding
``frequency'' $\Omega$ replaced by a new expression $\omega$, as
given in detail in Refs.~\cite{MS,WKB1}).
This improved WKB method applied to cosmological perturbations
with a linear turning point in $\omega$ does not however predict
amplitudes with a sufficient accuracy to lowest order~\cite{MS}.
We have shown in Ref.~\cite{WKB1} how the prediction
for the amplitude may be improved by using a next-to-leading WKB
approximation.
Our method involved an adiabatic expansion and the numerical evaluation
of the higher-order coefficients which yielded the spectra for a given
inflationary model (and were compared with the exact spectra of
power-law inflation~\cite{WKB1}).
\par
The power spectra to next-to-leading WKB order, in the
adiabatic expansion, are given by
\be
&&
{\cal P}_\zeta=
\frac{H^2}{\pi\,\epsilon_1\,m_{\rm Pl}^2}
\left(\frac{k}{a\,H}\right)^3
\frac{{\rm e}^{2\,\xi_{\rm II,S}}\,\left(1+g_{(1){\rm S}}^{\rm AD}\right)}
{\left(1-\epsilon_1\right)\,\omega_{\rm II,S}}
\nonumber
\\
\label{spectra_correct}
\\
&&
{\cal P}_h=
\frac{16\,H^2}{\pi\,m_{\rm Pl}^2}
\,\left(\frac{k}{a\,H}\right)^3\,
\frac{{\rm e}^{2\,\xi_{\rm II,T}}\,\left(1+g_{(1){\rm T}}^{\rm AD}\right)}
{\left(1-\epsilon_1\right)\,\omega_{\rm II,T}}
\ ,
\nonumber
\ee
where $m_{\rm Pl}$ is the Planck mass and all quantities are
evaluated in the super-horizon limit.
A crucial part of our method is the evaluation of $\xi_{II}$ and
$g_{(1)}^{\rm AD}$.
In particular, in this work 
%{\em Letter\/} 
we will estimate
{\em analytically\/} next-to-leading WKB corrections to
${\cal O}(\epsilon_i)$ and show that the leading WKB order
can also give predictions to ${\cal O}(\epsilon_i^2)$.
\par
For the precise definition of $g_{(1)}^{\rm AD}$ and its evaluation
we refer to Eq.~(61b) and Section~V of Ref.~\cite{WKB1}.
The quantity $\xi_{II}$ was also defined in Eq.~(26b) of the same
reference, and can in general be written as
\be
\xi_{\rm II}(\eta_{\rm f},\eta_0)=
\int_{\eta_{\rm f}}^{\eta_0}
\sqrt{A^2(\eta)-k^2\,\eta^2}\,\frac{\d\,\eta}{\eta}
\  ,
\ee
with $\eta_0$ the time for which the integrand vanishes,
$\eta_{\rm f}$ the super-horizon limit and $A^2(\eta)$
contains all the dependence on the horizon
flow functions $\epsilon_i(\eta)$.
In a manner similar to repeated integration by parts,
we can obtain general expressions valid for every $A^2(\eta)$
which contain terms easy to evaluate explicitly and new integrals
of sufficiently high order in the $\epsilon_i(\eta)$ so that
they can be neglected.
For example, on employing such a process once, we obtain
the identity
\be
\!\!\!\!
&&
\xi_{\rm II}(\eta_{\rm f},\eta_0)=
-\sqrt{A^2(\eta_{\rm f})-k^2\,\eta_{\rm f}^2}
\nonumber
\\
\!\!\!\!
&&
\ \
-\frac{A(\eta_{\rm f})}{2}
\,\ln\left[\frac{A(\eta_{\rm f})-\sqrt{A^2(\eta_{\rm f})-k^2\,\eta_{\rm f}^2}}
{A(\eta_{\rm f})+\sqrt{A^2(\eta_{\rm f})-k^2\,\eta_{\rm f}^2}}\right]
\nonumber
\\
\!\!\!\!
&&
\ \
-\int_{\eta_{\rm f}}^{\eta_0}
\frac{A'(\eta)}{2}
\,\ln\left[\frac{A(\eta)-\sqrt{A^2(\eta)-k^2\,\eta^2}}
{A(\eta)+\sqrt{A^2(\eta)-k^2\,\eta^2}}\right]
\d\,\eta
\ .
\label{xi_gen_exact}
\ee
Since the integral in the right~hand~side contains $A'$,
it is of (at least) one order higher in the $\epsilon_i$
than the remaining terms, which can be calculated explicitly.
More details, omitted here for the sake of brevity,
will be given in a forthcoming paper~\cite{sequel}.
\par
\noindent
{\em Next-to-leading~WKB~order and first~slow-roll~order.}
The adiabatic corrections $g_{(1)}^{\rm AD}$ in
Eqs.~(\ref{spectra_correct}) to leading order in the horizon
flow functions are given by 
\be
&&
g_{(1){\rm S}}^{\rm AD}
=\frac{37}{324}
-\frac{19}{243}\,
\left(\epsilon_1+\frac12\,\epsilon_2\right)
\nonumber
\\
\label{AD_corrections}
\\
&&
g_{(1){\rm T}}^{\rm AD}
=\frac{37}{324}-\frac{19}{243}\,\epsilon_1
\ .
\nonumber
\ee
We can now write the expressions for the scalar and tensor
spectra to next-to-leading WKB order
(indicated by the subscript WKB$*$) and first slow-roll order
(indicated by the superscript (1))
\begin{subequations}
\begin{widetext}
\be
&&
\mathcal{P}_{\zeta,\scriptscriptstyle{\rm WKB*}}^{(1)}=
\frac{H^2}{\pi\,\epsilon_1\,m_{\rm Pl}^2}\,
A_{\scriptscriptstyle{\rm WKB*}}\,
\left[1-2\,\left(D_{\scriptscriptstyle{\rm WKB*}}+1\right)\,\epsilon_1
-D_{\scriptscriptstyle{\rm WKB*}}\,\epsilon_2
-\left(2\,\epsilon_1+\epsilon_2\right)\,\ln\left(\frac{k}{k_*}\right)
\right]
\nonumber
\\
\label{P_SlowRoll_1}
\\
&&
\mathcal{P}_{h,\scriptscriptstyle{\rm WKB*}}^{(1)}=
\frac{16\,H^2}{\pi\,m_{\rm Pl}^2}\,
A_{\scriptscriptstyle{\rm WKB*}}\,
\left[1-2\,\left(D_{\scriptscriptstyle{\rm WKB*}}+1\right)\,\epsilon_1
-2\,\epsilon_1\,\ln\left(\frac{k}{k_*}\right)
\right]
\ ,
\nonumber
\ee
\end{widetext}
where $A_{\scriptscriptstyle{\rm WKB*}}=361/18\,e^3\approx 0.999$ and
$D_{\scriptscriptstyle{\rm WKB*}}\equiv\frac{7}{19}-\ln 3\approx -0.7302$.
%The coefficients $A^{(1)}$ and $D^{(1)}$ replace $A^{(0)}=18/e^3$ and
%$D^{(0)}=1/3 -\ln 3$ of the WKB approximation to leading order \cite{MS}.
The slow-roll approximation~\cite{SL} predicts, for the corresponding
quantities, $A_{\scriptscriptstyle{\rm SR}}=1$ and
$D_{\scriptscriptstyle{\rm SR}}=C\equiv\gamma_E+\ln 2-2\approx-0.7296$
(where $\gamma_E$ is the Euler-Mascheroni constant).
Thus, the next-to-leading WKB order gives an error of about $0.1\%$
for the estimate of the amplitude and one of about $0.08\%$ on the
coefficient $C$.
This shows that {\em analytical\/} results obtained by the WKB method
have reached the same accuracy as the standard slow-roll approximation.
We also obtain the same slow-roll spectral indices and
$\alpha$-runnings,
\be
&&
\!\!\!\!\!\!\!\!n_{\rm S,\scriptscriptstyle{\rm WKB*}}^{(1)}-1=
-2\,\epsilon_1-\epsilon_2
\ ,\quad
n_{\rm T,\scriptscriptstyle{\rm WKB*}}^{(1)}=-2\,\epsilon_1
\label{n_NTL_WKB}
\\
\nonumber
\\
&&
\!\!\!\!\!\!\!\!\alpha_{\rm S,\scriptscriptstyle{\rm WKB*}}^{(1)}=
\alpha_{\rm T,\scriptscriptstyle{\rm WKB*}}^{(1)}=0
\ ,
\label{alpha_NTL_WKB}
\ee
on using respectively Eqs.~(\ref{n_def}) and (\ref{alpha_def}).
From Eq.~(\ref{R_def}) the tensor-to-scalar ratio becomes
\be
R^{(1)}_{\scriptscriptstyle{\rm WKB*}}=
16\,\epsilon_1\left(1+D_{\scriptscriptstyle{\rm WKB*}}\,\epsilon_2\right)
\ .
\label{R_next-to-lead_WKB}
\ee
\end{subequations}
\par
\noindent
{\em Leading~WKB~order and second~slow-roll~order.}
We would now like to increase our accuracy in the slow-roll parameters
to second order, while keeping the WKB approximation to leading order
(a further improved treatment to second order in the slow-roll
parameters is in progress~\cite{sequel}).
On setting $g_{(1)}^{\rm AD}=0$ in Eqs.~(\ref{spectra_correct}),
we can write the expressions for the scalar and tensor perturbations
to leading~WKB~order (indicated by the subscript WKB),
and second~slow-roll~order (indicated with the superscript (2))
as
\begin{subequations}
\begin{widetext}
\be
\mathcal{P}_{\zeta,\scriptscriptstyle\scriptscriptstyle{\rm WKB}}^{(2)}
&\!\!\!=\!\!\!&
\frac{H^2}{\pi\,\epsilon_1\,m_{\rm Pl}^2}\,
A_{\scriptscriptstyle{\rm WKB}}
\left\{1-2\left(D_{\scriptscriptstyle{\rm WKB}}+1\right)\,\epsilon_1
-D_{\scriptscriptstyle{\rm WKB}}\,\epsilon_2
+\left(2\,D_{\scriptscriptstyle{\rm WKB}}^2
+2\,D_{\scriptscriptstyle{\rm WKB}}-\frac19\right)\,\epsilon_1^2
\right.
\nonumber
\\
&&\left.
+\left(D_{\scriptscriptstyle{\rm WKB}}^2-D_{\scriptscriptstyle{\rm WKB}}
+\frac{\pi^2}{12}-\frac{20}{9}\right)\,\epsilon_1\,\epsilon_2
+\left(\frac12\,D_{\scriptscriptstyle{\rm WKB}}^2+\frac29\right)\,\epsilon_2^2
+\left(-\frac12\,D_{\scriptscriptstyle{\rm WKB}}^2+\frac{\pi^2}{24}-
\frac{1}{18}\right)\,\epsilon_2\,\epsilon_3
\right.
\nonumber
\\
&&\left.
+\left[-2\,\epsilon_1-\epsilon_2
+2\left(2\,D_{\scriptscriptstyle{\rm WKB}}+1\right)\,\epsilon_1^2
+\left(2\,D_{\scriptscriptstyle{\rm WKB}}-1\right)\,\epsilon_1\,\epsilon_2
+D_{\scriptscriptstyle{\rm WKB}}\,\epsilon_2^2
-D_{\scriptscriptstyle{\rm WKB}}\,\epsilon_2\,\epsilon_3\right]\,
\ln\left(\frac{k}{k_*}\right)
\right.
\nonumber
\\
&&
\left.
+\frac12\,\left(4\,\epsilon_1^2+2\,\epsilon_1\,\epsilon_2+\epsilon_2^2
-\epsilon_2\,\epsilon_3\right)\,
\ln^2\left(\frac{k}{k_*}\right)
\right\}
\nonumber
\\
\label{P_SlowRoll_0_2order}
\\
\mathcal{P}_{h,\scriptscriptstyle{\rm WKB}}^{(2)}
&\!\!\!=\!\!\!&\frac{16\,H^2}{\pi\,m_{\rm Pl}^2}\,
A_{\scriptscriptstyle{\rm WKB}}
\left\{1-2\left(D_{\scriptscriptstyle{\rm WKB}}+1\right)\,\epsilon_1
+\left(2\,D_{\scriptscriptstyle{\rm WKB}}^2
+2\,D_{\scriptscriptstyle{\rm WKB}}-\frac19\right)\,\epsilon_1^2
\nonumber
\right.
\\
&&
\left.
+\left(-D_{\scriptscriptstyle{\rm WKB}}^2
-2\,D_{\scriptscriptstyle{\rm WKB}}+\frac{\pi^2}{12}-\frac{19}{9}\right)\,
\epsilon_1\,\epsilon_2
+\left[-2\,\epsilon_1+2\left(2\,D_{\scriptscriptstyle{\rm WKB}}+1\right)\,
\epsilon_1^2
-2\,\left(D_{\scriptscriptstyle{\rm WKB}}+1\right)\epsilon_1\,
\epsilon_2\right]\,
\ln\left(\frac{k}{k_*}\right)
\nonumber
\right.
\\
&&
\left.
+\frac12\,\left(4\,\epsilon_1^2
-2\,\epsilon_1\,\epsilon_2\right)\,\ln^2\left(\frac{k}{k_*}\right)
\right\}
\ ,
\nonumber
\nonumber
\ee
\end{widetext}
where $A_{\scriptscriptstyle{\rm WKB}}=18/e^3\approx 0.896$ and
$D_{\scriptscriptstyle{\rm WKB}}\equiv\frac{1}{3}-\ln 3\approx -0.7653$.
The spectral indices~(\ref{n_def}) are then given by
\be
n_{\rm S,\scriptscriptstyle{\rm WKB}}^{(2)}-1&=&
-2\,\epsilon_1-\epsilon_2-2\,\epsilon_1^2
-\left(2\,D_{\scriptscriptstyle{\rm WKB}}+3\right)\,\epsilon_1\,\epsilon_2
\nonumber
\\
&&
-D_{\scriptscriptstyle{\rm WKB}}\,\epsilon_2\,\epsilon_3
\label{n_2'order}
\\
n_{\rm T,\scriptscriptstyle{\rm WKB}}^{(2)}&=&
-2\,\epsilon_1-2\,\epsilon_1^2
-2\,\left(D_{\scriptscriptstyle{\rm WKB}}+1\right)\,\epsilon_1\,\epsilon_2
\nonumber
\ee
and their runnings~(\ref{alpha_def}) by
\be
\alpha_{\rm S,\scriptscriptstyle{\rm WKB}}^{(2)}=
-2\,\epsilon_1\,\epsilon_2
-\epsilon_2\,\epsilon_3
\ ,
\quad
\alpha_{\rm T,\scriptscriptstyle{\rm WKB}}^{(2)}=
-2\,\epsilon_1\,\epsilon_2
\label{alpha_2'order}
\ .
\ee
We note that the second order correction $-2\,\epsilon_1^2$
to $n_{\rm S}$ and $n_{\rm T}$ agrees with the exact spectral
index for power-law inflation~\footnote{For power-law
inflation $\epsilon_1=1/p$, $n_S=n_T$ and the exact spectral index is
$n_T=1-2/(p-1)$.
For large $p$, $n_T=1-2/p-2/p^2+{\cal O}(p^{-3})$.
See~\cite{LA} for the most recent result based on the
uniform approximation which contains the correct term 
$-2\,\epsilon_1^2$ in $n_T$.}.
The tensor-to-scalar ratio~(\ref{R_def}) becomes
\be
R^{(2)}_{\scriptscriptstyle{\rm WKB}}&=&
16\,\epsilon_1\left[1+D_{\scriptscriptstyle{\rm WKB}}\,\epsilon_2
+\left(D_{\scriptscriptstyle{\rm WKB}}+\frac19\right)\,\epsilon_1\,\epsilon_2
\right.
\nonumber
\\
&&
\phantom{16\,\epsilon_1}\ 
\left.
+\left(\frac12\,D_{\scriptscriptstyle{\rm WKB}}^2-\frac{2}{9}\right)\,\epsilon_2^2
\right.
\nonumber
\\
&&
\phantom{16\,\epsilon_1}\ 
\left.
+\left(\frac12\,D_{\scriptscriptstyle{\rm WKB}}^2-\frac{\pi^2}{24}
+\frac{1}{18}\right)\,\epsilon_2\,\epsilon_3
\right]
\ .
\label{R_lead_WKB}
\ee
\end{subequations}
%\end{widetext}
%
Of course, Eqs.~(\ref{P_SlowRoll_0_2order})--(\ref{alpha_2'order})
give the same results as obtained in Ref.~\cite{MS}, to first order
in the horizon flow functions. As already found in \cite{MS}, the leading 
WKB order gives an error of about $10\%$ for
the estimate of the amplitude and one of about $5\%$ on $C$ with
respect to standard slow-roll results.
The runnings $\alpha_{\rm S}$ and $\alpha_{\rm T}$ are predicted to
be of ${\cal O}(\epsilon_i^2)$~\cite{KT},
and in agreement with those obtained by the GFM~\cite{gongstewart,LLMS}.
It is notworthy that our power spectra and spectral
indices have the same polynomial structure in the $\epsilon_i$
as those of the latter references.
For the spectral indices $n_{\rm S}$ and $n_{\rm T}$ an
analogous structure is also confirmed by the uniform
approximation~\cite{HHJMP,HHHJMP,LA}~\footnote{An agreement on the
polynomial structure, but not in the numerical coefficients,
also holds for the spectral indices $n_{\rm S}$ and $n_{\rm T}$,
which are the present predictions of the uniform
approximation~\cite{HHJMP,HHHJMP,LA}.}.
\begin{figure}[t]
\includegraphics[width=0.5\textwidth]{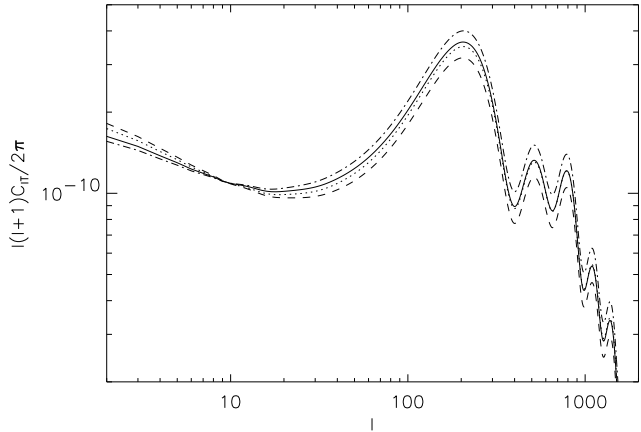}
\caption{
Comparison, for the same inflationary model
(with $\epsilon_1=\epsilon_2=0.1$,
$\epsilon_3=0$, $k_*=0.05$~Mpc$^{-1}$),
of angular power spectra for temperature anisotropies
of the cosmic microwave background due to scalar perturbations with different
accuracies in the spectral slope $n_{\rm S}$, as in Eqs.~(\ref{n_2'order})
and (\ref{alpha_2'order}).
The dotted line is the first-order slow-roll prediction, the dashed
line has $n_{\rm S}$ accurate to second order and no running,
the dot-dashed has $n_{\rm S}$ to first order and running
$\alpha_{\rm S}=-0.02$, the solid line is the full second order prediction.
The other parameters used for this flat model are
$\Omega_{\rm CDM}=0.26$, $\Omega_{b}=0.04$, $\Omega_{\Lambda}=0.7$ and
$H_0=72$ km s$^{-1}$ Mpc$^{-1}$.
Let us note that the WMAP first year observational relative error on the 
first peak is $0.7\%$ and on the second peak $1.2\%$~\cite{pageetal}.}
\label{Cl_grap}
\end{figure}
\par\noindent
{\em Conclusions.}
We have shown that inflationary theoretical predictions
have now reached expressions to second order in the slow-roll parameters
confirmed by two completely different approximation schemes, such as
the GFM and WKB.
As can be seen from Fig.~\ref{Cl_grap}, for some inflationary models, second
order slow-roll corrections to the power spectrum are necessary to perform
a correct comparison between theoretical predictions and observational
data.
The figure also shows that not only the runnings are important, but
also the ${\cal O}(\epsilon_i^2)$ terms in the spectral indices.
\par
The different predictions of the two methods for the numerical
coefficients in front of the ${\cal O}(\epsilon_i^2)$ terms are
at most of the order of $5\%$ for the spectral indices
and of $10\%$ in the amplitudes.
For slow-roll parameters $\epsilon_i\sim 0.1$, this leads
to an accuracy of about $0.5\%$ in the theoretical predictions for
the tensor-to-scalar ratio~(\ref{R_lead_WKB}).
A similar accuracy (of $0.1\%$) has also been reached by the Boltzmann 
codes~\cite{cmb}.
The predictions in the numerical coefficients and in the amplitude
can be further improved by employing the next-to-leading WKB
method~\cite{sequel}
while including terms to second order in the slow-roll parameters. 
\acknowledgments
We would like to thank Salman~Habib, Katrin~Heitmann and
Jerome~Martin for discussions and comments.

\end{document}